\documentclass[prd,preprint]{revtex4}
\usepackage{epsfig}

\begin{document}
\title{Collisional Dark Matter and Scalar Phantoms}
\author{D.\,E. Holz}
\email{deholz@itp.ucsb.edu}
\author{A. Zee}
\email{zee@itp.ucsb.edu}
\affiliation{Institute for Theoretical Physics, University of
California, Santa Barbara, CA 93106}


\begin{abstract}
As has been previously proposed, a minimal modification of
the standard $SU(3)\times SU(2)\times U(1)$ theory
provides a viable dark matter candidate. Such a particle, a
scalar gauge singlet, is naturally
self-interacting---making it of particular interest given
recent developments in astrophysics. We review this dark
matter candidate, with reference to the parameter ranges
currently under discussion.
\end{abstract}


\maketitle

\section{Introduction}

\label{sec:intro}
Although the presence of dark matter in the universe appears
to be no longer in doubt, the nature of the dark matter
remains elusive. Spergel and Steinhardt~\cite{KSS} have
suggested self-interacting dark matter as a possible
solution to numerous discrepancies between predictions and
observations in the standard dark matter scenario. Cold dark
matter predicts overly dense cores in galaxies, and they
argue that this can be resolved if the dark matter particles
interact with each other with a substantial scattering cross
section $\sigma$, while interacting only weakly with ordinary
matter. The basic physics is easy enough to understand:
collisional dark matter conducts heat, and as the core of a
galaxy heats up it expands, thus lowering the central
density. Furthermore, this form of dark matter may account
for the discrepancy between theory and observation on small
scale structure, the stability of galactic bars, and a
number of other issues.~\cite{wandelt}
In particular, Ostriker~\cite{ostriker} has shown that
collisional dark matter can also account for the growth rate
of massive black holes at the center of galaxies. For these
scenarios to work it is estimated that $\sigma /m,$ where
$m$ denotes the mass of the dark matter particle, must lie
in the range $0.5\mbox{--}6\mbox{ cm}^{2}/\mbox{g}$, or
equivalently, $2\times 10^{3}\mbox{--}3\times 10^{4}\mbox{
GeV}^{-3}$.


In this note we discuss a possible candidate within the
standard $SU(3)\times SU(2)\times U(1)$ theory. In all
likelihood, the standard theory descends from a theory at
higher energy scales, a grand unified theory, or possibly a
string theory. It seems plausible that in this descent there
would appear numerous scalar fields $X_{a}$ which happen to
be singlets under $SU(3)\times SU(2)\times U(1).$ These
fields would not couple to ordinary fermions, and would be
unknown to the standard gauge bosons. In general they can
couple to Higgs fields, and through the Higgs fields feebly
to ordinary fermions. The scalar field sector of the theory
might read
\begin{equation}
\frac{1}{2}\sum\limits_{a}m_{a}^{2}X_{a}^{2}+\frac{1}{4}\sum\limits_{a}\sum
\limits_{b}\eta _{ab}X_{a}^{2}X_{b}^{2}+\frac{1}{4}\sum\limits_{a}\sum
\limits_{j}\rho _{aj}X_{a}^{2}(\varphi _{j}^{\dagger }\varphi _{j}),
\end{equation}
where $\varphi _{j}$ denotes Higgs doublets, and $\eta _{ab}$
and $\rho _{aj}$ denote numerous coupling constants. The
$X_{a}^{\prime }$s interact with each other directly through
self interaction (the $\eta $ term), or indirectly
through Higgs exchange (generated by the $\rho $
term). Fifteen years ago, Silveira and one of us~\cite{old}
proposed these scalar particles, $X_{a}$ (called ``scalar
phantoms''), as candidates for the dark matter. We would
like to re-examine this possibility in light of recent
developments, including improvements in our knowledge of the
Higgs sector.

\section{Self-interacting dark matter candidate}

\label{sec:2}

Clearly the general case, with the set of parameters
$m_{a},$ $\eta _{ab}$ and $\rho _{aj}$, allows us a great
deal of freedom. Following Silveira and Zee, we go to the
simplest case of just one $X$ field and one Higgs doublet,
so that the relevant part of the Lagrangian reads
\begin{equation}
\frac{1}{2}m^{2}X^{2}+\frac{1}{4}\eta X^{4}+\frac{1}{4}\rho X^{2}(\varphi
^{\dagger }\varphi )-\frac{1}{2}\mu ^{2}\varphi ^{\dagger
}\varphi +\frac{1}{ 4}\lambda (\varphi ^{\dagger }\varphi
)^{2}. \label{model}
\end{equation}
We have imposed the discrete symmetry $X\rightarrow -X$,
which guarantees stability of the $X$ particle, as the $X$
field is odd under this symmetry while all other fields in
the universe are even. The coupling $\eta$ controls the
self-interaction of the $X$, which we will choose to be in
the range needed for cosmology. As we will see, this
simplest case may be barely viable.

After spontaneous symmetry breaking, $\varphi $ acquires a vacuum
expectation value $v$ with $v^{2}=\mu ^{2}/\lambda .$ Writing $|\varphi
(x)|=v+H(x)$ as usual, we have $m_{H}^{2}\sim \mu ^{2}$ and $m_{X}^{2}=m^{2}+
\frac{1}{2}\rho v^{2}.$ Replacing one of the $\varphi $ fields in the term $
\rho X^{2}(\varphi ^{\dagger }\varphi )$ by its vacuum expectation value, we
induce a cubic coupling of the form $\sim \rho vHX^{2}.$
Since the physical Higgs field $H$ couples to quark and
lepton fields, this cubic coupling leads
to a coupling of $X$ to quark and lepton fields. As $X$
couples to ordinary matter only via the Higgs field, it
naturally interacts feebly with ordinary matter and
could easily have escaped detection. Indeed, at low momentum
transfer $X$ interacts with a quark or lepton via the
exchange of the Higgs field, giving an interaction amplitude
$\sim \rho vf/m_{H}^{2}\sim \rho m_{f}/m_{H}^{2}.$ As usual,
$f$ denotes generically the Yukawa coupling of $H$ to a
quark or lepton and $m_{f}\sim fv$ denotes the mass of the
quark or lepton. This assumes that there is only one
$\varphi $ field; with multiple $\varphi $ fields these
estimates are all loosened. Since, on the scale of particle
physics, the masses $m_{f}$ of the quarks and leptons in the
first generation are rather small, the interaction of $X$
with ordinary matter is further suppressed by a factor $\sim
m_{f}/m_{X}$. In other words, as $X$ knows about ordinary
matter only through the Higgs, which itself couples very
weakly to the first generation of fermions, its interaction
with ordinary matter is necessarily weak, thus making $X$ a
natural candidate for the dark matter particle. (This also
means that in the early universe, when fermions of the
second and third generations were present in abundance, $X$
might have played a significant role.)

\bigskip

In its simplest version our model is governed by three
unknown parameters: $\eta $, $\rho $, and $m_{X}$. In the
absence of experimental input, Silveira and Zee~\cite{old}
made the natural choice that $\rho $ be roughly equal to
$\lambda $, the Higgs self coupling, but this choice is
certainly open to modification.

In the model of eq.~(\ref{model}) the scattering amplitude
for $ X+X\rightarrow X+X$ receives two contributions: $\sim
\eta $ from the direct quartic coupling, and $\sim (\rho
v)^{2}/m_{H}^{2}\sim (\rho ^{2}\mu ^{2}/\lambda
)/m_{H}^{2}\sim \rho ^{2}/\lambda $ from Higgs exchange$.$
The scattering amplitude is given by the larger of $\eta $
and $\rho ^{2}/\lambda .$ As we know nothing about $\eta $,
a natural assumption is to take $\eta $ and $\rho
^{2}/\lambda $ to be comparable, but again this is not
required. If we take the scattering amplitude for
$X+X\rightarrow X+X$ to be of order $\sim \rho ^{2}/\lambda
$, then at low momentum transfer the differential cross
section is given by $\sigma (X+X\rightarrow X+X)\sim (\rho
^{2}/\lambda )^{2}/m_{X}^{2}.$ Inserting $\sigma /m_{X}\sim
(\rho ^{2}/\lambda )^{2}/m_{X}^{3}$ into the
Spergel-Steinhardt bound ($2\times 10^{3}\mbox{ GeV}^{-3}\lesssim\sigma
/m_{X}\lesssim3\times 10^{4}\mbox{ GeV}^{-3}$) we obtain
\begin{equation}
m_{X}\sim 0.03\mbox{--}0.08\,(\rho ^{2}/\lambda
)^{2/3}\mbox{ GeV}.
\label{ss_bound}
\end{equation}
For example, for an $X$ particle with mass comparable to the nucleon we
would need $6\sqrt{\lambda }\lesssim \rho \lesssim 13\sqrt{\lambda }.$ This
parameter range is entirely reasonable by the standards of particle physics.
If, on the other hand, $\eta $ is larger than $\rho
^{2}/\lambda $, then eq.~(\ref{ss_bound}) is to be replaced by
\begin{equation}
m_{X}\sim 0.03\mbox{--}0.08\,\eta ^{2/3}\mbox{ GeV}.  \label{ss_bound2}
\end{equation}

\section{Coupling constants}

\label{sec:coupling}

As pointed out in the previous section, our proposed dark matter candidate
in the simplest version is governed by three parameters: the self-coupling, $
\eta $, the mass of the particle, $m_{X}$, and the coupling to $\varphi $,
given by $\rho $. Let us now study the limits set on their values by
cosmological considerations. The parameters $\eta $, $\rho ,$ and $m_{X}$
govern the temperature $T_{f}$ at which $X$ particles freeze out. This in
turn determines the number density at freeze out, and thus the current
number density and mass density of $X$ particles. By demanding that a
fraction of critical density $\Omega _{X}\sim 0.1$ is due to $X$ particles,
we are able to estimate the mass of the particle.

Following Lee \& Weinberg~\cite{LeeWeinberg},~we are to solve the evolution
equation
\begin{equation}
{\frac{dn}{dt}}=-{\frac{3\dot{R}}{R}}n-\langle \sigma v\rangle n^{2}+\langle
\sigma v\rangle n_{0}\/^{2},
\end{equation}
where $n$ denotes the number density of $X$ as a function of time $t$, $n_{0}
$ the equilibrium number density of $X$, $R$ the scale size of the Universe,
and $\langle \sigma v\rangle $ the thermally averaged cross section for
annihilation of two $X$ particles into fermions $X+X\to f+\bar{f}$. The
first term on the right-hand side describes the expansion of the Universe,
the second term accounts for particle annihilation, and the last term
represents particle production. We note that the contribution of
the cosmological constant to the critical density (estimated to be $\Omega 
_{\Lambda }\approx 0.7$ at present) stays constant as we go back in time,
while the matter density increases as $1/R^{3}$. At freeze
out, therefore, the cosmological constant is negligible.

The rate equation can be rewritten as
\begin{equation}
{\frac{df}{dx}}=\left( {\frac{45}{8\pi ^{3}N_{f}G}}\right)
^{1/2}m_{X}\langle \sigma v\rangle (f^{2}-f_{0}\/^{2}),
\end{equation}
where $x=T/m_{X}$, $f(x)=n/T^{3}$, $f_{0}(x)=n_{0}/T^{3}$,  $G$ is the
gravitational constant, and $N_{f}$ is the effective number of degrees of
freedom. Based on the numerical fit from Lee and Weinberg we obtain a
numerical approximation to the value of $f$ at freeze-out:
\begin{equation}
f(x_{f})\approx 10\left[ {\left( {\frac{45}{8\pi ^{3}N_{f}G}}\right)
^{1/2}m_{X}\langle \sigma v\rangle }\right] ^{-0.95}.  \label{num_soln}
\end{equation}

The number density of $X$ particles today is given by $n_0=f(x_f) T_0\/^3$,
where $T_0=2.7\ \mbox{K}=2.4\times10^{-13}\ \mbox{GeV}$ is the current
temperature. The present mass density in $X$ is given by $\rho_X=m_X n_0$,
and so the contribution to closure density is
\begin{equation}
\Omega_X = f(x_f) T_0\/^3 m_X/\bar\rho,  \label{closure}
\end{equation}
with $\bar\rho$ the present day critical density needed to close the
Universe.

The freeze out temperature, $T_{f}=x_{f}m_{X}$, is given by
\begin{equation}
{\frac{1}{\sqrt{x_{f}}}}e^{1/x_{f}}=\left( {\frac{45}{16\pi ^{6}N_{f}G}}
\right) ^{1/2}m_{X}\langle \sigma v\rangle .  \label{freeze}
\end{equation}

\section{Numbers}

\label{sec:numbers}

Even in the simple model of eq.~(\ref{model}) there is a
great deal of freedom in setting values for the different
parameters. For the sake of definiteness, we take the
annihilation cross section $\langle \sigma v\rangle $ to
be~\cite{old}
\begin{equation}
\langle \sigma v\rangle \approx {\frac{3(\rho m_{f})^{2}}{4\pi 
(4m_{X}\/^{2}-m_{H}\/^{2})^{2}}},
\end{equation}
where $m_{H}$ is the mass of the Higgs, $\rho $ is a coupling constant (from
the $X^{2}\varphi ^{+}\varphi $ term in the Lagrangian), and $m_{f}$ is the
mass of the heaviest fermion lighter than $X$.

Approximating the exponent in eq.~(\ref{num_soln}) as $-1$, and plugging
this into eq.~(\ref{closure}), we find that the explicit mass dependence
drops out,~\cite{KolbTurner} and
\begin{eqnarray}
\Omega_X&\sim&10 \sqrt{\frac{8\pi^3 N_f G}{45}} {\frac{T_0\/^3}{\bar\rho}} {
\frac{1}{\langle\sigma v\rangle}} \\
&\sim&{\frac{10^{-10}\sqrt{N_f}}{h^2 \langle\sigma v\rangle}}\ 
\mbox{millibarn},
\end{eqnarray}
with the Hubble constant given by $H_0= 100\,h\ \mbox{km
s}^{-1}\mbox{Mpc}^{-1}$. Utilizing our expression for the cross section, the
contribution of $X$ to the closure density comes out to be
\begin{equation}
\Omega_X\sim 10^{-9} {\frac{\sqrt{N_f}}{h^2}} {\frac{(4 m_X\/^2-m_H\/^2)^2}{
(\rho m_f)^2}}\mbox{ GeV}^{-2}.
\end{equation}
We can assume that $\rho$ takes a value of order $1$. For simplicity, we
take $\rho\sim{\frac{1}{8}}g^2(m_H/m_W)^2$,~\cite{old} with $g$ the weak
coupling constant and $m_W=80\ \mbox{GeV}$ the mass of the W. For the range
of parameters of interest to us, the heaviest appropriate fermion will be
the bottom quark, so $m_f=m_b=4\ \mbox{GeV}$. Taking $h\sim0.7$, $N_f\sim10$
, and demanding that $\Omega_X\sim0.3$, gives a mass of
\begin{equation}
m_X\sim m_H/2.  \label{mass}
\end{equation}

Plugging these values into eq.~(\ref{freeze}), we find that the freezing
temperature is given by $x_f=T_f/m\sim0.04$ (which depends only weakly on $
m_X$). The particle is thus non-relativistic when it freezes out, as has
been assumed throughout.

It is clear that a parameter range can be found for $\rho$
and $m_X$ such that the $X$ can make a significant
contribution to the dark matter in the Universe. However,
from eqs.~(\ref{ss_bound}) and~(\ref{ss_bound2}), our mass
range implies high values for the coupling constants,
suggesting that our perturbative approach is breaking down,
and that our calculations are only to be taken
heuristically. The simple case considered here, consisting
of a single $X$ field and a single Higgs doublet, is only
marginally viable.  As remarked earlier, the general version
of our model allows us a great deal of freedom, and it is
likely that trivial generalizations of this simple case could
lead to viable dark matter candidates.

\bigskip
\noindent
Note added: In the process of writing this note we learned
that similar ideas were discussed by M. C. Bento,
O. Bertolami, R. Rosenfeld, and L.  Teodoro (Phys. Rev. D
{\bf 62}, 041302(R) (2000)), and also by C. P. Burgess,
M. Pospelov, and T. Veldhuis (hep-ph/0011335).

\acknowledgments
AZ would like to thank Jerry Ostriker and Paul Steinhardt
for informative discussions. DEH acknowledges very useful
conversations with Dave Spergel, Mark Srednicki, and Neil
Turok. This work was supported in part by the NSF under
grant number PHY 99-07949 at the ITP.




\end{document}